\documentclass[aip,reprint,nofootinbib,superscriptaddress]{revtex4-1}

\pdfoutput=1

\usepackage{graphicx}
\usepackage{dcolumn}
\usepackage{bm}
\usepackage{amsmath}
\usepackage{amssymb}
\usepackage{braket}
\usepackage{booktabs}
\usepackage{multirow}
\usepackage{setspace}
\usepackage{enumitem}
\usepackage{xr}
\usepackage{pdfpages}

\makeatletter
\AtBeginDocument{\let\LS@rot\@undefined}
\makeatother

\begin{document}

\title{Engineering Nonlinear Response of Superconducting Niobium Microstrip Resonators via Aluminum Cladding}

\author{Sangil Kwon}
\email{kwon2866@gmail.com}
\affiliation{Institute for Quantum Computing, University of Waterloo, Waterloo, Ontario N2L 3G1, Canada}
\affiliation{Department of Physics and Astronomy, University of Waterloo, Waterloo, Ontario N2L 3G1, Canada}

\author{Yong-Chao Tang}
\affiliation{Institute for Quantum Computing, University of Waterloo, Waterloo, Ontario N2L 3G1, Canada}
\affiliation{Department of Electrical and Computer Engineering, University of Waterloo, Waterloo, Ontario N2L 3G1, Canada}

\author{Hamid~R.~Mohebbi}
\affiliation{High Q Technologies LP, Waterloo, Ontario N2L 0A7, Canada}

\author{Olaf~W.~B.~Benningshof}
\affiliation{Institute for Quantum Computing, University of Waterloo, Waterloo, Ontario N2L 3G1, Canada}
\affiliation{Department of Physics and Astronomy, University of Waterloo, Waterloo, Ontario N2L 3G1, Canada}

\author{David G. Cory}
\affiliation{Institute for Quantum Computing, University of Waterloo, Waterloo, Ontario N2L 3G1, Canada}
\affiliation{Department of Chemistry, University of Waterloo, Waterloo, Ontario N2L 3G1, Canada}
\affiliation{Perimeter Institute for Theoretical Physics, Waterloo, Ontario N2L 2Y5, Canada}
\affiliation{Canada Institute for Advanced Research, Toronto, Ontario M5G 1Z8, Canada}

\author{Guo-Xing Miao}
\affiliation{Institute for Quantum Computing, University of Waterloo, Waterloo, Ontario N2L 3G1, Canada}
\affiliation{Department of Electrical and Computer Engineering, University of Waterloo, Waterloo, Ontario N2L 3G1, Canada}

\date{\today}

\begin{abstract}
In this work, we find that Al cladding on Nb microstrip resonators is an efficient way to suppress nonlinear responses induced by local Joule heating, resulting in improved microwave power handling capability.
This improvement is likely due to the proximity effect between the Al and the Nb layers.
The proximity effect is found to be controllable by tuning the thickness of the Al layer.
We show that improving the film quality is also helpful as it enhances the microwave critical current density, but it cannot eliminate the local heating.
\end{abstract}

\maketitle

\section{Introduction}
\label{sec:intro}

Recently, superconducting planar resonators \cite{zmuidzinas} have found applications in magnetic resonance because their low dissipation and small mode volume have greatly improved the sensitivity of spin detection \cite{xiang, kurizki, benningshof2013, malissa2013, sigillito2014, grezes2014, bienfait2016a, eichler2017, sigillito2017, probst2017}.
In addition to the detection sensitivity, superconducting resonators are required to handle strong microwave pulses for efficient spin manipulation.
However, microwave power handling capability of superconducting planar resonators has been an issue---if we apply a strong microwave pulse, the nonlinearity of the resonator participates such that the actual microwave magnetic field that a spin sees becomes significantly different from what we intended.

Regarding the Duffing-type nonlinearity \cite{duffing}, it is known that we can design a pulse which can compensate the nonlinear response as the Duffing-type nonlinearity can be easily modeled and controlled using nonlinear circuit models \cite{oates1993, tholen2009, ku2010, swenson2013, mohebbi2014, hincks2015}.
However, many reported nonlinear responses of type-II superconducting resonators are very difficult to model, thus not controllable: the shape of S-parameter curves becomes irregular and greatly suppressed even at modest microwave power.
This type of nonlinearity has been attributed to local Joule heating, often called a hot spot\cite{gurevich1987}, followed by switching of weak links to the normal state \cite{jacobs1996, wosik1997, wosik1999, purnell2004, abdo2006, wosik2009, ghigo2009, ghigo2010, ku2010, brenner2011, kurter2011}.
Thus, it is crucial to minimize this undesired nonlinear response for magnetic resonance applications.

In this work, we investigated ways to suppress the nonlinearity due to local Joule heating by improving the film quality (Sec.~\ref{sec:nonlinear_noAl}) and Al cladding (Sec.~\ref{sec:nonlinear_Al}) \cite{tang2017}.
We first showed that local Joule heating is the dominant source of the nonlinearity of pure Nb microstrip resonators.
Then, we found that a resonator made of better quality film showed a significantly higher microwave critical current, but the major mechanism of the nonlinearity remains the same.
Meanwhile, Al cladding effectively eliminated the nonlinear responses induced by local Joule heating.
This improved microwave power handling capability is likely due to the proximity effect between the Al and the Nb layers \cite{deGennes, jin, vissers2013, oxford1}.
The existence of the proximity effect was confirmed experimentally by studying how magnetic field dependence of the resonance frequency $f$ and the quality factor $Q$ change as we tune the thickness of the Al layer (Sec.~\ref{sec:lossPara}). This study also showed that the proximity effect is controllable by tuning the thickness of the Al layer.

\section{Methods}
\label{sec:method}

\begin{table}
\caption{The name convention of resonators and their resonance frequency $f_0$ and loaded quality factor $Q_0$ below 20 mK in zero-field.
The first column introduces each resonator's name used in this paper.
The second column shows the composition of microstrips' heterostructure.
Numbers indicate the thickness of each layer in nm.
$T_\textrm{growth}$ is the substrate temperature when the films for microstrips were grown.
The exact values of $T_\textrm{growth}$ and detailed film growth conditions are shown in Table~\ref{tab:growth}.
}
\label{tab:device}\centering
\begin{ruledtabular}
\begin{tabular}{l c c c c}
\noalign{\smallskip}
Res.		&	Composition	&	$T_\textrm{growth}$	&	$f_0$ (GHz)	&	$Q_0$ \\
\noalign{\smallskip} \hline \noalign{\smallskip}
Al-0L	&	Nb$\,$50	&	low	&	10.0728	&	14300 \\
Al-5L	&	Al$\,$5/Nb$\,$50/Al$\,$5	&	low	&	\hphantom{1}9.9764	&	21000 \\
Al-10L	&	Al$\,$10/Nb$\,$50/Al$\,$10	&	low	&	10.0672	&	23600 \\
Al-0H	&	Nb$\,$50	&	high	&	10.0792	&	27500 \\
\end{tabular}
\end{ruledtabular}
\end{table}

\begin{table*}
\caption{Summary of the Nb film growth conditions. The two rightmost columns are resulting critical temperature $T_\textrm{c}$ and residual resistivity $\rho_\textrm{n}$ from transport measurements. RT stands for Room Temperature. ``Ar sputter cleaning'' means Ar sputter cleaning of the substrate, i.e., back sputtering, before the film growth. For further details about the growth conditions, see Ref.~\onlinecite{tang2017} and the supplementary material of Ref.~\onlinecite{kwon}, in which the films for Al-0H and Al-0L are appeared as wafers A and C, respectively. For more details about the resulting film properties, see Table I of Ref.~\onlinecite{kwon}.}
\label{tab:growth}\centering
\begin{ruledtabular}
\begin{tabular}{l c c c c c c c c c}
\noalign{\smallskip}
Res.	&	\begin{tabular}{@{}c@{}}Base pressure \\ (mbar)\end{tabular}	&	\begin{tabular}{@{}c@{}} Growth rate \\ (\AA/s)\end{tabular}	&	\begin{tabular}{@{}c@{}} Ar pressure \\ (mbar)\end{tabular}	&	\begin{tabular}{@{}c@{}}Temperature: \\ strip ($^\circ$C)\end{tabular}	&	\begin{tabular}{@{}c@{}}Temperature: \\ ground plane ($^\circ$C)\end{tabular}	&	\begin{tabular}{@{}c@{}}Ar-sputter \\ cleaning\end{tabular}	&	\begin{tabular}{@{}c@{}}$T_\textrm{c}$ \\ (K)\end{tabular}	&	\begin{tabular}{@{}c@{}}$\rho_\textrm{n}$ \\ ($\mu\Omega$ cm)\end{tabular} 	\\
\noalign{\smallskip} \hline \noalign{\smallskip} \noalign{\smallskip}
Al-0L	&	$\sim\,$10$^{-8\hphantom{1}}$	&	0.6		&	$4 \times 10^{-3}$	&	RT	&	RT	&	Yes	&	7.2	&	17		\\
Al-0H	&	$\sim\,$10$^{-10}$	&	1.7		&	$2 \times 10^{-3}$	&	550	&	770	&	No	&	9.3	&	2.9		\\
\end{tabular}
\end{ruledtabular}
\end{table*}

Four microstrip resonators with different film quality and Al cladding thickness were used.
Two of them were pure Nb resonators with different film quality.
The film quality was controlled mainly by the temperature of the substrate at the time of the film growth;
the higher substrate temperature resulted in higher critical temperature and lower residual resistivity, i.e., a better quality film.
The other two resonators were trilayer resonators---Nb resonators with Al cladding.
For all resonators, the ground plane was made of a pure Nb layer, and the thickness of Nb layers, including the ground plane, was 50 nm.
Since our resonators are all microstrip resonators, i.e., double-sided film, there is no complication for making the ground plane and the microstrip with different material compositions.

The resonators are labeled with the thickness of the Al layers and the substrate temperature as shown in Table~\ref{tab:device};
Al-5L and Al-10L indicate that the thickness of each layer is Al/Nb/Al = 5/50/5 nm and 10/50/10 nm, respectively, and the films were grown at low temperature.
A control sample of pure Nb grown at higher temperature is labeled Al-0H.
The detailed growth conditions of the Nb films are summarized in Table~\ref{tab:growth}.
The trilayer resonators were grown in one vacuum cycle with the same condition as that of Al-0L to avoid potential alloying at elevated temperature.
High-resolution XRD pattern of the trilayer film can be found in Ref.~\onlinecite{tang2017}.
All films were grown by DC magnetron sputtering on both sides of $c$-plane sapphire wafers (double-side-polished 430 $\mu$m thick and 2 inch diameter).
Then the resonators were fabricated by optical lithography and dry etching.

\begin{figure}
\centering
\includegraphics[scale=0.5]{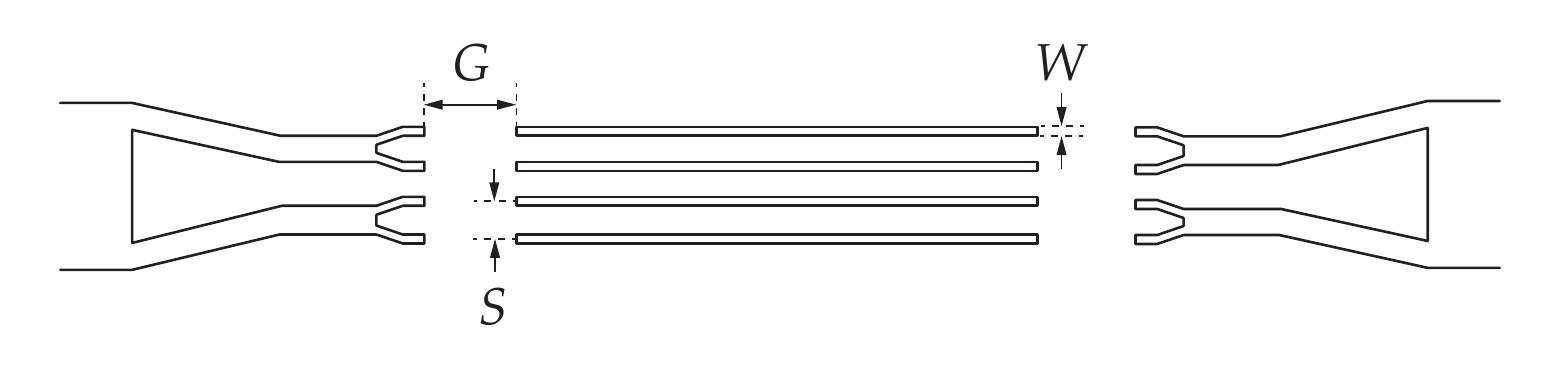}
\caption{\label{fig:config}
Geometry of the resonators.
$G$ is the gap between the feedline and the resonator. $W$ is the width of a strip. $S$ is the spacing between center of strips.
The value of $G$ is 400 for Al-0H and 350 $\mu$m for other resonators; 
$W$, 15 $\mu$m; and $S$, 75 $\mu$m.
The distance between the strips and the ground plane is 430 $\mu$m.
For clarity, the ground plane is not shown.
}
\end{figure}

The resonators are straight half-wavelength resonators;
each resonator was composed of four microstrips as shown in Fig.~\ref{fig:config} (Ref.~\onlinecite{mohebbi2014}).
The geometry of all resonators are identical, except the coupling gap ($G$ in Fig.~\ref{fig:config}).
For Al-0H, the coupling gap was 400 $\mu$m;
for other resonators, the coupling gap was 350 $\mu$m.
The resonance frequencies of all resonators below 20 mK without a magnetic field are about 10 GHz.

The reason for employing microstrip resonators is that this geometry can generate uniform microwave fields above the strips compared to coplanar resonators.\cite{benningshof2013}
Therefore, we believe a microstrip geometry is more suitable for electron spin resonance (ESR) of thin films, which is our research interest.\cite{mohebbi2014, kwon}
As a trade-off, the less confined field profile inherently leads to more radiation loss and dielectric loss;
thus, our resonators show lower internal quality factor (see Table~\ref{tab:fit}) values than corresponding coplanar resonators.

The measurements were made in a cryogen-free dilution refrigerator (Leiden CF250).
The resonator is aligned with a magnetic field using a goniometer (Attocube ANGt101) with the precision $\pm 5$ mdeg at 0.1 K.
For more details, see Sec.~III of Ref.~\onlinecite{kwon}.

Full S-parameters were collected using a vector network analyzer (VNA, Agilent N5230A).
The resonance frequency and the loaded quality factor $Q_\textrm{load}$ were obtained by fitting the magnitude of the measured $S_{21}$ to a complex Lorentzian function \cite{kwon}.
The external quality factor $Q_\textrm{ex}$ was obtained using the formula $Q_\textrm{load} = Q_\textrm{ex} 10^{-\textrm{IL}/20}$, where IL is the insertion loss in dB (Ref.~\onlinecite{sage2011}).
The internal quality factor $Q_\textrm{in}$ was obtained from the relation $Q_\textrm{load}^{-1} = Q_\textrm{ex}^{-1} + Q_\textrm{in}^{-1}$.
The input microwave power $P_\textrm{in}$ was estimated by $P_\textrm{in} = P_\textrm{VNA} - \textrm{CL}$ (in dB), where $P_\textrm{VNA}$ is the setting power of the VNA, and CL is the loss in cables from the VNA to the input capacitor of the resonator.\footnote{The estimated insertion loss can vary 1--2 dB from one package to the next mainly due to imperfect package assembling. Such a deviation is not crucial for the power dependence, but it may change the values of the internal quality factor in Table~\ref{tab:fit} about 20--30\%.}
The circulating power $P_\textrm{circ}$ was estimated using $P_\textrm{circ} = \pi^{-1} P_\textrm{in} Q_\textrm{load} 10^{-\textrm{IL}/20}$.
The maximum value of the microwave current $I_\textrm{mw}$ circulating in the resonator was estimated using $I_\textrm{mw} = \sqrt{8 P_\textrm{circ} / Z_0}$ (Ref.~\onlinecite{oates1990}),
where $Z_0$ is the characteristic impedance of the resonator, which is designed to be about 50 $\Omega$.

\section{Improving the Film Quality}
\label{sec:nonlinear_noAl}

\begin{figure*}
\centering
\includegraphics[scale=0.5]{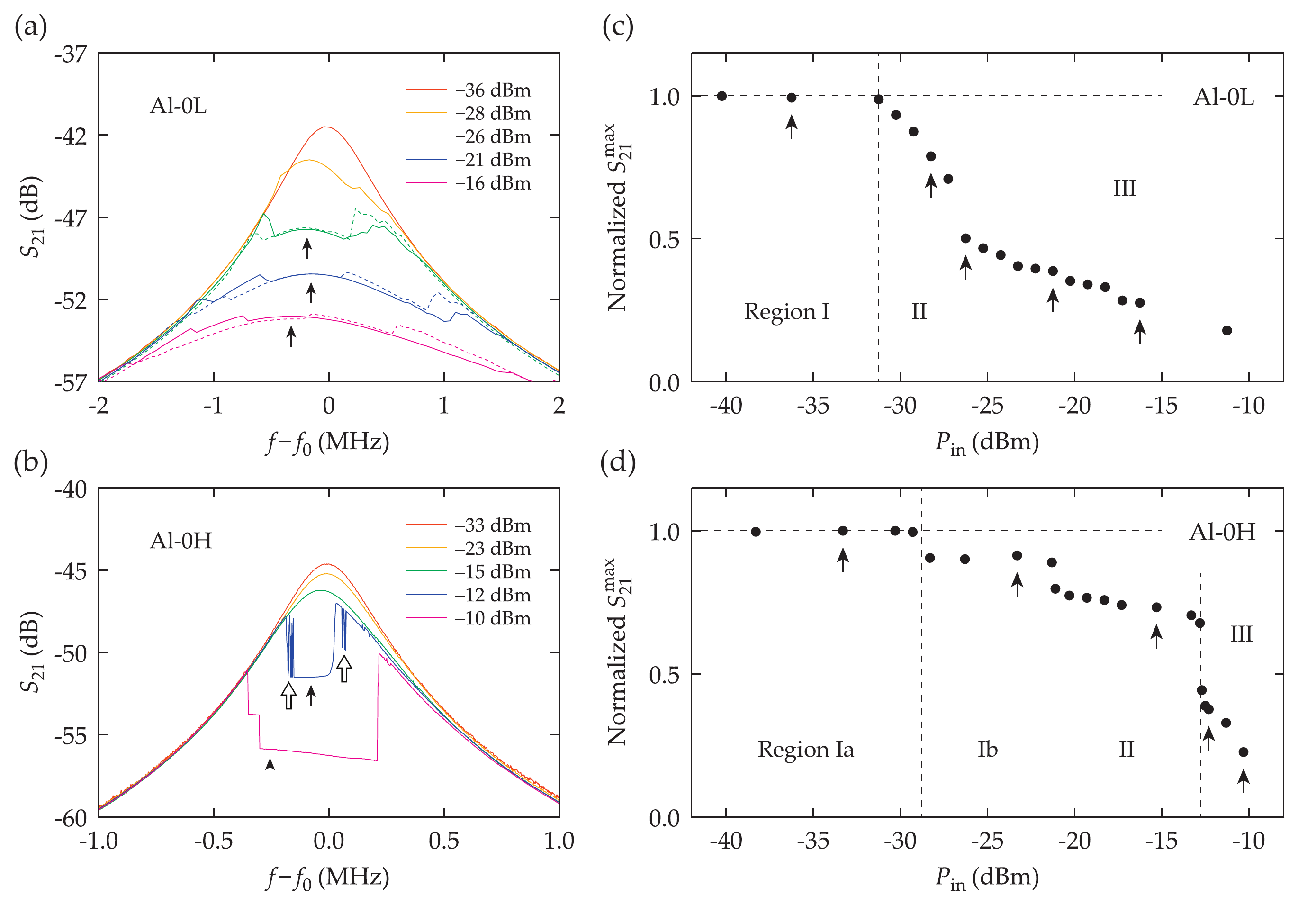}
\caption{\label{fig:power_noAl} 
(a,b) $S_{21}$ resonance curves of the pure Nb resonators at various $P_\textrm{in}$.
Solid lines were swept from low to high frequency; dashed lines in (a) were swept in the opposite direction.
Hollow arrows in (b) denote bistable behaviors.
(c,d) The maximum value of $S_{21}$ for each curve (in linear scale), $S_{21}^\textrm{max}$, as a function of $P_\textrm{in}$.
Region I is the low-power region at which the $S_{21}$ curve is Lorentzian.
In region II, the curve becomes asymmetric and distorted.
In region III, the onset of jump in $S_{21}^\textrm{max}$ appears and the shape of the $S_{21}$ curve becomes very irregular.
The data were normalized by the value of $S_{21}^\textrm{max}$ at the lowest $P_\textrm{in}$.
$P_\textrm{in}$ for $S_{21}$ curves in (a,b) are denoted by arrows.
In (d), to address an abrupt drop in $S_{21}^\textrm{max}$ and the quality factor while maintaining the Lorentzian shape and the same resonane frequency, we divide region I into regions Ia and Ib.
In region III, $S_{21}^\textrm{max}$ is defined by the maximum $S_{21}$ in the crator, as denoted by arrows in (a,b).
The data were taken at zero field and about 0.2 K.
}
\end{figure*}

Figure~\ref{fig:power_noAl}(a) shows $S_{21}$ curves of Al-0L at various $P_\textrm{in}$.
As $P_\textrm{in}$ increases, the resonance peak becomes rapidly suppressed and irregular.
To know the source of this nonlinearity, we plotted the maximum value of $S_{21}$ ($S_{21}^\textrm{max}$) as a function of $P_\textrm{in}$ [Fig.~\ref{fig:power_noAl}(c)].
Note that the plot can be divided into three regions based on the shape of the $S_{21}$ curve:
In the low-power region (region I), the $S_{21}$ curve is Lorentzian.
As $P_\textrm{in}$ increases (region II), the curve becomes asymmetric and distorted.
Finally, the resonator enters region III with the onset of jump in $S_{21}^\textrm{max}$.
In this region, the shape of the $S_{21}$ curve is very irregular;
when the curves were swept in the opposite direction, the $S_{21}$ curve becomes reflected shape [dashed lines in Fig.~\ref{fig:power_noAl}(a)].

The same measurements were done with a resonator made of a better quality Nb film, Al-0H [Fig.~\ref{fig:power_noAl}(b)].
There are several notable differences between Al-0L and Al-0H:
First, for Al-0H, the shape of the $S_{21}$ curve is Lorentzian and the resonance frequency remains the same until $P_\textrm{in}$ reaches $-21$ dBm.
However, there is an abrupt drop in $S_{21}^\textrm{max}$ and the quality factor near $P_\textrm{in} = -29$ dBm; hence, we divide region I into regions Ia and Ib.
The reason for this sudden drop is unclear at this stage.

Secondly, in region III, Al-0H shows cratered Lorentzian shapes and bistability, which have been accounted for switching of weak links, such as grain boundaries, to the normal state \cite{wosik1997, wosik1999, purnell2004, abdo2006, wosik2009, ghigo2009, ghigo2010, ku2010, brenner2011, kurter2011, gurevich1987, hylton1988, hylton1989, halbritter1990, halbritter1992, halbritter1995}.
For Al-0L, the irregular shape is reproducible from sweep to sweep---there is no notable bistable behavior.
This suggests that the spread in the microwave critical currents of Al-0L more significant than that of Al-0H such that bistable behaviors are averaged out, resulting in spike-like features (see Fig.~9 in Ref.~\onlinecite{ghigo2010}).

\begin{table}
\caption{
Circulating powers ($P_\textrm{circ}$) and microwave current densities ($j_\textrm{mw}$) near the outermost edges at the boundaries between the regions in Fig.~\ref{fig:power_noAl}(c,d).
The superscript indicates the boundary;
for example, the quantity at the boundary between regions I and II have the superscript I,II.
The DC depairing current density predicted by the Ginzburg--Landau theory $j_\textrm{d}^\textrm{GL}$ is also shown.
$j_\textrm{mw}^\textrm{I,II}$ and $j_\textrm{mw}^\textrm{II,III}$ were obtained by $I_\textrm{mw}$  (see Sec.~\ref{sec:method}) and simulations for the current density distribution.
$j_\textrm{d}^\textrm{GL}$ was calculated using $j_\textrm{d}^\textrm{GL} = (2/3)^{1.5} H_\textrm{c}/\lambda$ (Ref.~\onlinecite{matsushita}).
For Al-0L, $\mu_0 H_\textrm{c}$ and $\lambda$ were taken from Table~\ref{tab:fit};
for Al-0H, 0.27 T and 52 nm, respectively \cite{kwon}.
The units are dBm for circulating powers and A/m$^2$ for current densities.
}
\label{tab:currentDensity}\centering
\begin{ruledtabular}
\begin{tabular}{l c c c c c}
\noalign{\smallskip}
Res.	&	$P_\textrm{circ}^\textrm{I,II}$	&	$P_\textrm{circ}^\textrm{II,III}$	&	$j_\textrm{mw}^\textrm{I,II}$	&	$j_\textrm{mw}^\textrm{II,III}$	&	$j_\textrm{d}^\textrm{GL}$	\\
\noalign{\smallskip} \hline \noalign{\smallskip} \noalign{\smallskip}
Al-0L	&	1.2	&	4	&	$1.5 \times 10^{10}$	&	$2 \times 10^{10}$	&	$5 \times 10^{11}$	\\
Al-0H	&	11	&	18	&	$1.3 \times 10^{11}$	&	$3 \times 10^{11}$	&	$2 \times 10^{12}$	\\
\end{tabular}
\end{ruledtabular}
\end{table}

For quantitative understanding, we estimated microwave current densities at the boundaries between regions; the values are summarized in Table~\ref{tab:currentDensity}.\footnote{To obtain $j_\textrm{mw}^\textrm{II,III}$, we needed $Q$ factors of the data in region II. Since the shape of $S_{21}$ curves is slightly asymmetric, the Lorentzian fitting was not perfect. We just used the values from the fitting because the asymmetry was not significant. We also tried a 3 dB bandwidth and it gave similar values.}
Here, note that, $j_\textrm{mw}^\textrm{II,III}$, which can be considered as the microwave critical current density, is about one order of magnitude less than $j_\textrm{d}^\textrm{GL}$ for both resonators, suggesting that the microwave power handling capability is not limited by intrinsic and global properties of superconductivity.
This supports that the dominant mechanism for the nonlinearity of pure Nb films is local Joule heating.
Also note that $j_\textrm{mw}^\textrm{II,III}$ of the resonator made of a better quality film (Al-0H) are about one order of magnitude higher than that of the resonator with low film quality (Al-0L).
This shows that improving the film quality enhances the microwave critical current density, but it does not change the dominant mechanism for the nonlinearity.

\section{Aluminum Cladding}
\label{sec:nonlinear_Al}

\begin{figure}
\centering
\includegraphics[scale=0.5]{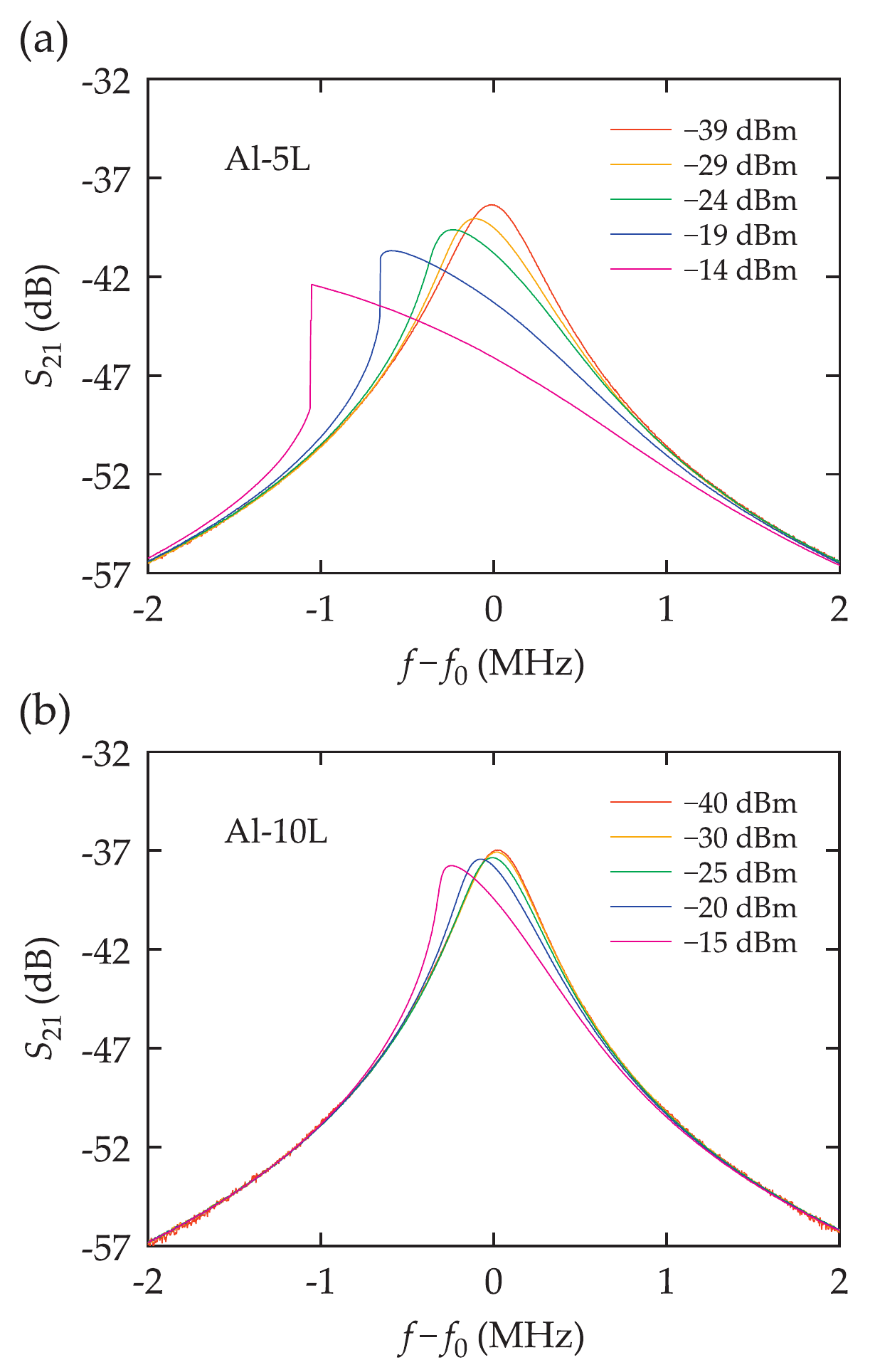}
\caption{\label{fig:power_Al} 
$S_{21}$ resonance curves of the trilayer resonators at various $P_\textrm{in}$.
The sweep direction was from low to high frequency.
The data were taken at zero field.
The data were taken at at zero field and about 0.2 K.
}
\end{figure}

Figure~\ref{fig:power_Al} shows $S_{21}$ curves at various $P_\textrm{in}$ after Al cladding.
Note that Al cladding changes the nonlinear response dramatically: Al-5L and Al-10L show the Duffing-type nonlinearity instead of irregular shapes.
Since this type of nonlinearity is controllable using nonlinear circuit models \cite{oates1993, ku2010, swenson2013, mohebbi2014, hincks2015}, as mentioned in Sec.~\ref{sec:intro}, we can say that Al cladding improves the high-power handling capability.

The qualitative change in the nonlinearity after Al cladding is likely due to current bypasses provided by Al near weak links in the Nb layer.
The superconductivity of these Al bypasses is strengthened by the proximity effect such that the critical current density and the critical field of the Al layers are substantially enhanced.
Here, to provide reliable bypasses, the Al layers need to cover the surface of the Nb layers perfectly.
In this regard, Al and Nb combination is special because Al grown at room temperature wets on the surface of Nb ideally\cite{chang1987}. The reason for this is that the bonding between Al and Nb is stronger than that between Al and Al.

Other possible roles of Al cladding, such as protection against oxidation of the Nb layer \cite{halbritter1987, halbritter1988} and enhancing the thermal conductivity \cite{feshchenko2017}, can reduce the number of weak links and local Joule heating;
however, they cannot fully account for such a qualitative change.
Here, note that a normal metal layer can do the same things, except strengthening the superconductivity.
In Refs.~\onlinecite{ghigo2009, ghigo2010}, a 35 nm thick layer of Au was deposited on an MgB$_2$ thin film, but the nonlinear behavior due to switching of weak links remained largely unchanged.
In addition, another way to reducing the number of weak links, improving the film quality, does not change the dominant mechanism for the nonlinearity as shown in Sec.~\ref{sec:nonlinear_noAl}.
From these results, we believe that the qualitative change in nonlinear response after Al cladding is mainly due to the proximity effect.

Even after the switching of weak links are eliminated, further Al cladding can still assist, as shown in Fig.~\ref{fig:power_Al}(b).
Note that, in Table~\ref{tab:fit}, Al-10L shows higher internal quality factor ($Q_\textrm{0,in}$), which implies that Al cladding reduces the surface resistance, and lower effective residual resistivity ($\rho_\textrm{n,fit}$) compared to Al-5L. (See Sec.~\ref{sec:lossPara} for further explanation regarding Table~\ref{tab:fit}.)
These results suggest that the observed nonlinearity of the trilayer resonators is driven by global heating\cite{wosik1999, wosik2009} which is generated through the following process:
When the circulating power is low enough such that there is no notable nonlinear response, the heat balance between the cooling power and the dissipated power due to the finite surface resistance is fulfilled.
At this stage, there are not many thermally-excited quasiparticles because of the low temperature (0.2 K).
When the circulating power passes a certain level, at which the dissipated power is greater than the cooling power, quasiparticles are excited and participate in the power dissipation, which is proportional to $\rho_\textrm{n} j_\textrm{mw}^2$, where $\rho_\textrm{n}$ is the residual resistivity \cite{kwon}.

The intrinsic GL nonlinearity\cite{gittleman1965, lam1992, clem2012} does not account for the nonlinearity of the trilayer resonators because the GL nonlinearity is known to be much more reactive than shown by the data in Fig.~\ref{fig:power_Al} (Ref.~\onlinecite{golosovsky1995}).
Indeed the data in Fig.~\ref{fig:paraField} of Sec.~\ref{sec:lossPara}, which follow the GL equations closely, the shift of $f$ about 1 MHz does not result in notable change in $Q$ and $S_{21}^\textrm{max}$.
In addition, vortex penetration into grains is also unlikely because, if vortices were created by a microwave current and penetrated into the grains, a hysteretic behavior would be observed due to vortex pinning;
in other words, the $S_{21}$ curve would not go back to its original position and shape once high $P_\textrm{in}$ was applied \cite{ghigo2009}.
Such behavior was not observed at zero field for all resonators.

In a modest field parallel to the microwave current $H_\parallel$, we find that the results in Figs.~\ref{fig:power_noAl} and \ref{fig:power_Al} are largely unchanged. Some of the representative data are shown in Fig.~S3.
In a field perpendicular to the film $H_\perp$, we found that applying a high microwave current results in magnetic hysteresis caused by suppression of the edge barrier and consequent injection of vortices.
Here, these vortices are created by $H_\perp$, not by the microwave current.
The supporting data and analysis are in Sec.~S3.

\section{The Proximity Effect}
\label{sec:lossPara}

\begin{figure}
\centering
\includegraphics[scale=0.5]{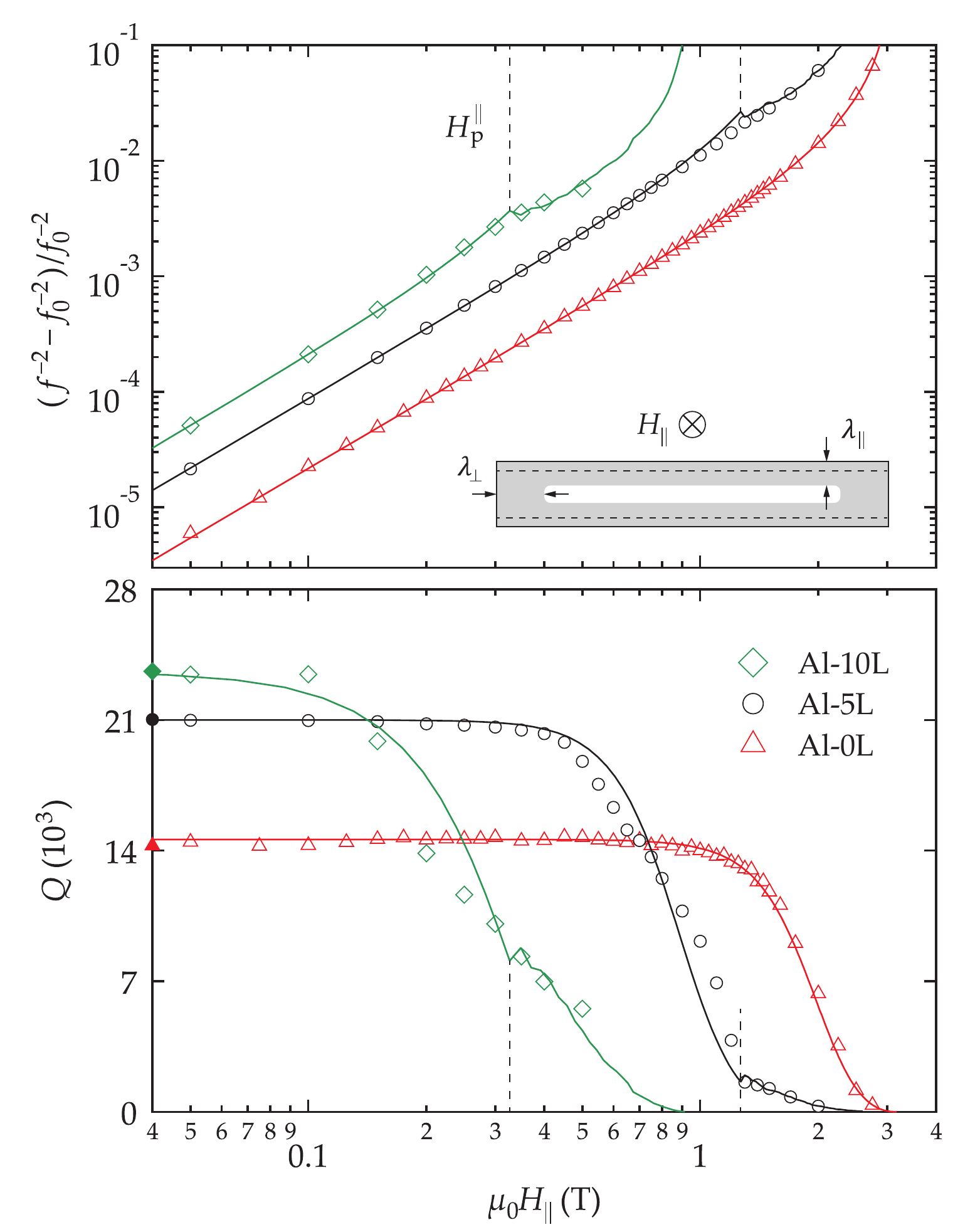}
\caption{\label{fig:paraField} 
Parallel magnetic field $H_\parallel$ dependence of $f^{-2}$ and $Q$ after zero-field cooling, where $f$ and $Q$ are the resonance frequency and the loaded quality factor.
Solid lines are from calculations with parameters in Table~\ref{tab:fit}.
The anisotropy parameter $\gamma$ is assumed 1.
The vertical dashed lines indicate the vortex penetration field parallel to the film $H_\textrm{p}^\parallel$ obtained from the solutions of the anisotropic GL equations.
Solid symbols in the lower panel represent zero-field loaded quality factor.
Al-0L data are from Ref.~\onlinecite{kwon}.
The mixing chamber temperature was below 20 mK.
The circulating power was kept about $-20$ dBm to avoid any nonlinear response.
Errors are comparable to or smaller than the size of symbols.
The inset shows the pictorial definition of in-plane penetration depth $\lambda_\parallel$ and out-of-plane penetration depth $\lambda_\perp$ in the cross-sectional view when there is no vortex.
The gray area is the region penetrated by an external magnetic field.
Here, $\lambda_\parallel$ is assumed to be isotropic.
Dashed lines are interfaces between the Al and Nb layers.
Note that the figure in the inset is not in scale;
in our experimental configuration, $H_\parallel$ almost completely penetrates the film because the film thickness is less than $\lambda_\parallel$ (see Table~\ref{tab:fit}).
}
\end{figure}

\begin{table*}
\caption{Loss parameters extracted from Fig.~\ref{fig:paraField} by following the procedures described in Sec.~S2.
The internal quality factor below 20 mK without a magnetic field $Q_\textrm{0,in}$ and the external quality factor $Q_\textrm{ex}$ are also shown.
Here, $Q_\textrm{ex}$ of the resonators in this table are almost the same because the gap between the feedline and the resonator ($G$ in Fig.~\ref{fig:config}) are designed to be identical for straightforward comparison of the quality factors.
$\lambda_0^\parallel$ is the zero-field in-plane penetration depth;
$\gamma$ is the anisotropy parameter;
$\kappa_\parallel$ is the in-plane GL parameter;
$H_\textrm{c}$ is the thermodynamic critical field;
$Q_\textrm{0,fit}$ is the zero-field loaded quality factor determined by fitting;
$\beta$ is the exponent for the fraction of normal electrons in the context of the two-fluid picture (see Sec.~S2 for the formal definition); and
$\rho_\textrm{n,fit}$ is the residual resistivity obtained from fitting. 
The loss parameters of Al-0L are from Ref.~\onlinecite{kwon}.
Note that, given the data in Fig.~\ref{fig:paraField}, $\kappa_\parallel$ and $\gamma$ cannot be determined independently;
any combination of $\kappa_\parallel$ and $\gamma$ gives similar results if $\gamma\kappa_\parallel$ is the same.
The reason is that $H_\textrm{vp}^\parallel$ is roughly proportional to $\xi_\parallel \xi_\perp$ (Ref.~\onlinecite{matsushita}), and $\xi_\parallel \xi_\perp$ is proportional to $\gamma \kappa_\parallel$.
}
\label{tab:fit}\centering
\begin{ruledtabular}
\begin{tabular}{l c c c c c c c c}
\noalign{\smallskip}
Res.	&	$Q_\textrm{ex}$	&	$Q_\textrm{0,in}$	&	$\lambda_0^\parallel$ (nm)	&	$\gamma\kappa_\parallel$	&	$\mu_0 H_\textrm{c}$ (mT)	&	$Q_\textrm{0,fit}$	&	$\beta$	&	$\rho_\textrm{n,fit}$ ($\mu\Omega \cdot$cm)	\\
\noalign{\smallskip} \hline \noalign{\smallskip} \noalign{\smallskip}
Al-10L	&	$4 \times 10^4$	&	$6 \times 10^4$	&	85		&	5.3	&	77	&	$2.36 \times 10^4$	&	1.1	&	4.3	\\
Al-5L	&	$4 \times 10^4$	&	$5 \times 10^4$	&	200		&	12.2	&	102	&	$2.10 \times 10^4$	&	2.2	&	14	\\
Al-0L	&	$4 \times 10^4$	&	$2 \times 10^4$	&	162		&	6.5	&	190	&	$1.46 \times 10^4$	&	2.2	&	17	\\
\end{tabular}
\end{ruledtabular}
\end{table*}

In this section, we experimentally confirm the existence of the proximity effect between the Al and the Nb layers and show this effect is controllable by varying the thickness of the Al layer.
For this, we investigate how a magnetic field parallel to the microwave current $H_\parallel$ dependence of $f$ and $Q$ change as we tune the thickness of the Al layer.
Figure~\ref{fig:paraField} shows how the thickness of the Al layers affects the $H_\parallel$ dependence of $Q$.
Note that, as the thickness of the Al layers increases, $Q$ starts to drop at a lower field, which already indicates the existence of the proximity effect.

In addition, as already mentioned in Sec.~\ref{sec:nonlinear_Al}, $Q_\textrm{0,in}$ becomes higher as the thickness of the Al layers increases.
The origin of higher $Q_\textrm{0,in}$ after Al cladding is probably less-lossy surface oxide or an improved interface between the substrate and the film.
Thermal quasiparticles are not relevant because, at the measurement temperature ($\lesssim\,$20 mK), thermal quasiparticles are expected to be frozen out.

For quantitative analysis of the proximity effect, we characterize the $H_\parallel$ dependence of $f^{-2}$ and $Q$ in Fig.~\ref{fig:paraField} using a set of parameters, which we call loss parameters, associated with magnetic field induced quasiparticle generation.
The basis of this approach is that the magnetic field dependence of the real and imaginary parts of the complex resistivity $\rho_1 + \textrm{i}\rho_2$ can be studied via $Q$ and $f$ as a function of field, respectively \cite{kwon}.
We emphasize that these loss parameters, shown in Table~\ref{tab:fit}, were obtained purely by comparing the measured and expected $f$ and $Q$ without incorporating any other types of measurements.
To calculate the expected $f$ and $Q$ as a function of $H_\parallel$, we need to model the complex resistivity \cite{kwon}.

In order to model the complex resistivity associated with quasiparticle generation, the anisotropic GL equations were used because the trilayer resonators are anisotropic systems.
(See Sec.~S1 for details on the implementation of the anisotropic GL equations.)
In this case, the penetration depth must be defined based on the direction of the magnetic field penetration as shown in the inset of Fig.~\ref{fig:paraField}.
The GL coherence lengths along the in-plane $\xi_\parallel$ and out-of-plane $\xi_\perp$ also need to be distinguished. Consequently, we have two GL parameters, $\kappa_\parallel$($\equiv\! \lambda_\parallel / \xi_\parallel$) and $\kappa_\perp$($\equiv\! \lambda_\perp / \xi_\perp$).
Here, $\gamma$ provides the relation $\lambda_\perp = \gamma \lambda_\parallel$.
Once the anisotropic GL equations were implemented, the expected $f$ and $Q$ as a function of $H_\parallel$ were calculated following the procedure described in Sec.~S2.
During the calculation, we treated the trilayer as a single anisotropic layer with the thickness of the whole trilayer.
Therefore, the loss parameters of Al-5L and Al-10L are effective parameters.

The measured data and calculated curves agree well (Fig.~\ref{fig:paraField}).
This suggests that our system can be treated as a single system with effective parameters, and we should not strongly separate Nb and Al layers in our system.
The reason for this is that the Al thickness is well below the coherence length of both Nb and Al.
If there is a phase transition from superconducting Al/superconducting Nb to normal Al/superconducting Nb, then there must be some abrupt change in the $H_\parallel$ dependence other than vortex injection, but no such a signature was observed.

In Table~\ref{tab:fit}, $H_\textrm{c}$ and $\rho_\textrm{n,fit}$ decrease as the Al thickness increases.
This result is consistent with previous reports \cite{zhao1999, wang2000, brammertz2001, brammertz2002}.
Note that $\lambda_0^\parallel$ of Al-5L is longer than that of Al-0L, although the proximity effect is expected to reduce $\lambda_0^\parallel$ (Refs.~\onlinecite{zhao1999, wang2000}).
This elongation of $\lambda_0^\parallel$ is likely due to electron scattering at the interface \cite{wang2000}.
As the Al layer becomes thicker, the contribution of the Al layer to $\lambda_0^\parallel$ becomes dominant compared with the interface. As a result, $\lambda_0^\parallel$ of Al-10L is significantly shorter than that of Al-5L.

Note that, in Table~\ref{tab:device}, $f_0$ of Al-5L is about 0.1 GHz lower than that of Al-0L. In addition, $f_0$ of Al-10L is 0.1 GHz higher than that of Al-5L.
Since $f_0$ is proportional to $1/\sqrt{L}$, where $L$ is the effective inductance per unit length, 0.1 GHz change in $f_0$ means 2\% change in $L$.
In our geometry, the dominant contribution to $L$ is the inductance from the energy stored as an electromagnetic field $L_\textrm{field}$.
From the simulation\cite{kwon}, we obtain $L_\textrm{field} \approx 510$ nH/m, suggesting that 0.1 GHz change in $f_0$ corresponds to about 10 nH/m change in $L$.
The kinetic inductance per unit length $L_\textrm{KI}$ can be calculated using the following formula\cite{kwon, schmidt}:
\begin{equation}
L_\textrm{KI} \approx \mu_0 \lambda^2 \int \frac{|J_\textrm{mw}|^2}{|I|^2} dA,
\end{equation}
where $A$ is the cross-sectional area of the resonator.
Using $\lambda$ in Table~\ref{tab:fit} and $J_\textrm{mw}$ from the simulation, we obtain the following values of $L_\textrm{KI}$:
10 nH/m for Al-0L; 13 nH/m for Al-5L; and 2.5 nH/m for Al-10L.
These results suggest that the difference in $f_0$ of Al-5L and Al-10L is due to the reduction of the kinetic inductance, i.e., the penetration depth, by thicker Al cladding.
However, the difference in $f_0$ of Al-0L and Al-5L is not easy to understand.
The Al-Nb interface might contribute to the inductance, but the mechanism is unclear.

Lastly, from Fig.~\ref{fig:paraField}, we found that 5 nm Al-cladding is a good choice for applications in X-band ESR of $g\,$$=\,$2 electron spin systems, which require a magnetic field of about 0.35 T.
However, if the film thickness or the film quality of the Nb layers is significantly different from that of Al-0L, then the optimal thickness of the Al layer may vary.

\section{Conclusion}

In conclusion, we found that nonlinear responses of pure Nb microstrip resonators were induced by local Joule heating, while that of Al-clad resonators was induced by global heating.
This qualitative change in nonlinear responses was likely due to Al current bypasses whose superconductivity is strengthened by the proximity effect between the Al and the Nb layers.
This proximity effect was found to be controllable by tuning the Al layer thickness: as the thickness of the Al layer increases, $\lambda_0$, $H_\textrm{c}$, and $\rho_\textrm{n}$ decrease.
Improving the film quality enhanced the microwave critical current density, but it did not result in a qualitative change in nonlinear responses.
Thus, our study showed that Al cladding is an effective way to eliminate nonlinear responses induced by local Joule heating, resulting in improved microwave power handling capability.

Strong microwave power handling capability will allow us to control spins or solid-state qubits efficiently\cite{samkharadze2016, kroll2019}. Hence, this work will be useful for magnetic resonance applications as well as quantum information processing.

\section*{Supplementary Material}

See the supplementary material for details regarding solving the anisotropic GL equations (Sec.~S1), extracting the loss parameters (Sec.~S2), and magnetic hysteresis in a finite $H_\perp$ (Sec.~S3). $S_{21}$ curves at various $P_\textrm{in}$ in a modest $H_\parallel$ are shown in Fig.~S3.

\begin{acknowledgements}
S.K. thanks to Ivar A. J. Taminiau and George Nichols for technical support, and Shinyoung Lee for helpful discussions.
This work is supported by the Canada First Research Excellence Fund, 
the Canada Excellence Research Chairs (grant No. 215284), 
the Natural Sciences and Engineering Research Council of Canada (grant Nos. RGPIN-418579 and RGPIN-04178), 
and the Province of Ontario.
The University of Waterloo's Quantum NanoFab was used for this work. 
This infrastructure is supported by
the Canada Foundation for Innovation, 
the Ontario Ministry of Research \& Innovation, 
Industry Canada, and 
Mike \& Ophelia Lazaridis. 
\end{acknowledgements}

\newpage
$ $
\includepdf[pages=1]{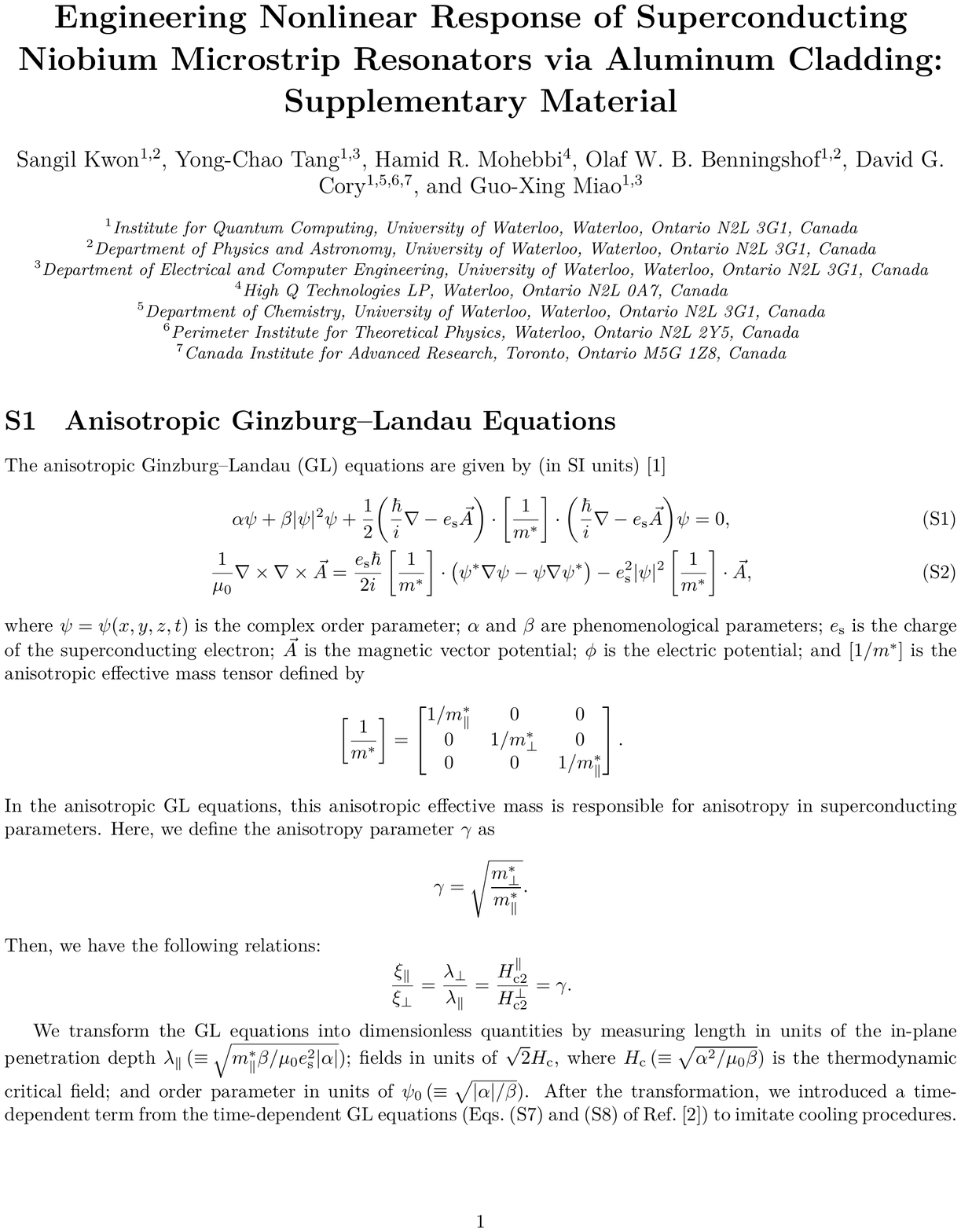}
\newpage
$ $
\includepdf[pages=2]{supp.pdf}
\newpage
$ $
\includepdf[pages=3]{supp.pdf}
\newpage
$ $
\includepdf[pages=4]{supp.pdf}
\newpage
$ $
\includepdf[pages=5]{supp.pdf}
\newpage
$ $
\includepdf[pages=6]{supp.pdf}

\end{document}